\documentclass[english,aps,prb]{revtex4}
\usepackage[T1]{fontenc}
\usepackage[latin9]{inputenc}
\usepackage{units}
\usepackage{bm}
\usepackage{graphicx}
\usepackage{esint}

\makeatletter
\@ifundefined{textcolor}{}
{%
 \definecolor{BLACK}{gray}{0}
 \definecolor{WHITE}{gray}{1}
 \definecolor{RED}{rgb}{1,0,0}
 \definecolor{GREEN}{rgb}{0,1,0}
 \definecolor{BLUE}{rgb}{0,0,1}
 \definecolor{CYAN}{cmyk}{1,0,0,0}
 \definecolor{MAGENTA}{cmyk}{0,1,0,0}
 \definecolor{YELLOW}{cmyk}{0,0,1,0}
}

\makeatother

\usepackage{babel}
\begin{document}

\title{Ion-acoustic solitons in warm magnetoplasmas with super-thermal electrons}

\author{$^{1}$B. Choudhury, $^{1}$R. Goswami, $^{2}$G. C. Das, and $^{1}$M.
P. Bora}

\email{mpbora@gauhati.ac.in}

\selectlanguage{english}%

\affiliation{$^{1}$Department of Physics, Gauhati University, Guwahati-781014,
India\\
$^{2}$Material Sceinece Division, Institute of Advanced Studies in
Science \& Technology,  Guwahati-781035, India}
\begin{abstract}
In this work, the phenomenon of formation of localised electrostatic
waves (ESW) or soliton is considered in a warm magnetoplasma with
the possibility of non-thermal electron distribution. The parameter
regime considered here is relevant in case of magnetospheric plasmas.
We show that deviation from a usual relaxed Maxwellian distribution
of the electron population has a significant bearing in the allowed
parameter regime, where these ESWs can be found. We further consider
the presence of more than one electron temperature, which is inspired
by recent space-based observations\cite{key-2}.
\end{abstract}
\maketitle

\section{Introduction}

Electrostatic solitary waves (ESW) or solitons are common occurrences
in the near-earth plasmas and are routinely observed in the boundary
layers and turbulent regions of the magnetosphere \cite{key-21}.
In recent years, many authors have tried to model these ESWs with
the help of different physical models \cite{key-2-1,key-3-1,key-4-1,key-5-1,key-6-1,key-7-1}.
Most of the observational and theoretical studies have indicated that
these ESWs are basically potential structures (compressive and rarefactive)
with density structures (enhancement and compression) and weak double
layers \cite{key-21}. Solitons or double layers are plasma sheaths
(discontinuities) moving in a plasma, which require competing species
of charged particles with different masses (inertia) and charges.
In an electron-ion plasma, these requirements are fulfilled due to
colder ions with their large inertia, which helps in formation of
plasma sheaths. So, an electron sheath in a plasma can be effectively
modified by equally hot ions \cite{key-9-1}. 

However, majority of these theoretical analyses do not consider the
effect of magnetic field into account. A full-blown analysis of solitary
waves where magnetic perturbation is self-consistently considered
can be very rigorous \cite{key-1}. Several space-borne experiments
have observed ESWs moving parallel to the ambient magnetic field in
various near-earth plasma environments such as the solar wind, magnetosheath
and magnetotail regions, and auroral zone \cite{key-2,key-3,key-4,key-5}.
These ESWs are, in general, bipolar structures moving in the direction
of the background magnetic field \cite{key-6}. In the auroral region,
these ESWs are reported as negative potential structures travelling
upward along the auroral magnetic field lines \cite{key-7,key-8}.
These observations are also supported by those from the Freja satellite
data \cite{key-9,key-10}. Besides, space plasmas, in general, can
be largely modeled with Maxwellian velocity distribution. However,
advancement of satellite based technologies in recent years, has led
to the realization that most of these plasmas, especially the near-earth
plasmas, have high energy tails and heat-flux shoulders, which may
be attributed to the fact that these plasmas are quite inhomogeneous
and semi-collisionless \cite{key-11,key-12,key-13}. Subsequently,
it has been established that these plasmas are best modeled by a generalized
Lorentzian or kappa distribution (especially the electron distributions)
rather than by a pure Maxwellian \cite{key-14,key-15}. Experimental
observations on solar wind plasmas, in recent years, have established
that these plasmas have a spectral index \cite{key-16} $\sim2.8$.
In the Earth's magnetosphere, the spectral index is typically observed
\cite{key-17} in the range $2\leq\kappa\leq8$. We, in this Chapter,
consider these ESWs in the presence of a super thermal electron component.

We, in this work have considered formation of ion-acoustic electrostatic
solitary waves in the presence of a background magnetic field. We
also consider the electron polulation to be super-thermal, modeled
through a Lorentzian or kappa velocity distribution. We further consider
the effect of two species of electrons with different temperatures.
Presence of multi-temperature electrons in magnetosphere is reported
experimentally \cite{key-2}. In Section I, we present the plasma
model that we have considered for formation of these ESWs in a warm
magnetoplasma. In Section II, we consider the solitary wave solutions
of the model, where we have incorporated the effect of super-thermal
electrons. In Section III, we consider two independent components
of super-thermal electrons with two different temperatures. We however
have shown that the two-temperature electrons have only marginal effect
on the structure and parameter regimes of the ESWs. Finally, we summarise
our conclusions in Section IV. The parameter regime considered in
this work and the related results can be relevant in explaining large-amplitude
ESWs observed in the earth's magnetosphere.

\section{Basic model of plasma}

Below we write down the basic governing equations for a thermal plasma,
which is immersed in an external magnetic field $\bm{B}_{0}$,

\begin{eqnarray}
\frac{\partial n}{\partial t}+\nabla\cdot(n\bm{v}) & = & 0,\label{eq:continuity}\\
\frac{d\bm{v}}{dt} & = & -\frac{e}{m_{i}}\nabla\phi-\frac{1}{m_{i}}\bm{\nabla}p+(\bm{v}\times\bm{\Omega}),\label{eq:momentum}\\
n=n_{e} & = & F(\phi),\label{eq:boltzmann}\\
p & \propto & (nm_{i})^{\gamma},\label{eq:state}
\end{eqnarray}
where Eqs.(\ref{eq:continuity}) and (\ref{eq:momentum}) represent
the continuity equation and conservation of momentum for the ions.
The function $F(\phi)$ represents equilibrium electron (and ion)
density according to the particular velocity distribution function
(VDF) i.e. Boltzmanian or kappa distribution. The last equation is
the equation of state. In the above equations, $n$ and $n_{e}$ are
the ion and electron densities and $\bm{\Omega}=e\bm{B}_{0}/(m_{i}c)$
is the ion gyro-frequency. Other symbols have their usual meanings
and quasi-neutrality is assumed all throughout. We assume the external
magnetic field to be in the $\hat{\bm{z}}$ direction.

We assume an arbitrary electrostatic perturbation in time and space
and define a co-moving coordinate $\eta=l_{x}x+l_{z}z-v_{M}t$, where
$l_{x,z}$ are direction cosines and thus defined by the relation
$l_{x}^{2}+l_{z}^{2}=1$, and $v_{M}$ is the velocity of the nonlinear
wave. Far away from the perturbation we assume everything to be stationary
and define the boundary condition as at $\eta\rightarrow\infty,n\rightarrow n_{0},\phi\rightarrow0,$
and $\bm{v}\rightarrow0$. Without any loss of generality, we can
assume that the physical quantities are constant along $\hat{\bm{y}}$
direction.

\subsection{Reduction of equations}

We now describe a general procedure for reducing Eqs.(\ref{eq:continuity}-\ref{eq:state}).
From the continuity equation, we can write
\begin{equation}
l_{x}v_{x}+l_{z}v_{z}-v_{M}=-v_{M}\frac{n_{0}}{n}.\label{eq:eq1}
\end{equation}
From the $x,y,$ and $z$ components of the momentum equation, we
can write,
\begin{eqnarray}
-v_{M}\frac{n_{0}}{n}v_{x}' & = & -l_{x}[f(n)+g(n)]+v_{y}\Omega,\label{eq:eq3}\\
-v_{M}\frac{n_{0}}{n}v_{y}' & = & -v_{x}\Omega,\label{eq:eq2}\\
-v_{M}\frac{n_{0}}{n}v_{z}' & = & -l_{z}[f(n)+g(n)],\label{eq:eq4}
\end{eqnarray}
where the $(')$ refers to derivative with respect to the scaled coordinate
$\eta$, and $f(n)$ and $g(n)$ functions of $n$,
\begin{eqnarray}
f(n) & = & \frac{T}{m_{i}(n_{0}^{2}n)^{1/3}}\, n',\\
g(n) & = & \frac{e}{m_{i}}\,\phi'.
\end{eqnarray}
Here, we have expressed the equilibrium presurre $\gamma p_{0}/n_{0}=T$,
with $\gamma$ being the ratio of the specific heats taken as $5/3$.
Note that all throughout the calculations, temperature $T$ is expressed
in energy units. We now take a derivative of Eqs.(\ref{eq:eq3}) and
(\ref{eq:eq4}) with respect to $\eta$ to get the following relations,
\begin{eqnarray}
-v_{M}\frac{n_{0}}{n}\left(v_{x}''-\frac{n'}{n}v_{x}'\right) & = & -l_{x}[f'(n)+g'(n)]+v_{x}\frac{n}{n_{0}}\frac{\Omega^{2}}{v_{M}},\label{eq:eq6}\\
-v_{M}\frac{n_{0}}{n}\left(v_{z}''-\frac{n'}{n}v_{z}'\right) & = & -l_{z}[f'(n)+g'(n)],\label{eq:eq5}
\end{eqnarray}
where we have substituted for $v_{y}'$ from Eq.(\ref{eq:eq2}). By
differentiating Eq.(\ref{eq:eq1}), successively with respect to $\eta$,
we get,
\begin{eqnarray}
l_{x}v_{x}'+l_{z}v_{z}' & = & h(n)\equiv v_{M}\frac{n_{0}}{n^{2}}n',\label{eq:eq7}\\
l_{x}v_{x}''+l_{z}v_{z}'' & = & p(n)\equiv-2v_{M}\frac{n_{0}}{n^{3}}(n')^{2}-v_{M}\frac{n_{0}}{n^{2}}n''.\label{eq:eq8}
\end{eqnarray}
Using Eqs.(\ref{eq:eq6}-\ref{eq:eq8}), we get,
\begin{equation}
-v_{M}\frac{n_{0}}{n}\left[p(n)-\frac{n'}{n}h(n)\right]=-[f'(n)+g'(n)]+\Omega^{2}\left(\frac{n}{n_{0}}-1\right)-l_{z}v_{z}\frac{n}{n_{0}}\frac{\Omega^{2}}{v_{M}},\label{eq:eq10}
\end{equation}
where we have used the condition $l_{x}^{2}+l_{z}^{2}=1$ and substituted
for $l_{x}v_{x}$ from Eq.(\ref{eq:eq1}). Eq.(\ref{eq:eq4}) can
be integrated to have,
\begin{equation}
v_{z}=\frac{l_{z}}{v_{M}n_{0}}\int n[f(n)+g(n)]\, d\eta+C_{1}\equiv q(n),\label{eq:vz}
\end{equation}
where $C_{1}$ is the constant of integration, to be evaluated by
imposing the boundary conditions. So, finally, using Eq.(\ref{eq:vz}),
from Eq.(\ref{eq:eq10}), we arrive at a single nonlinear, second
order differential equation for the system as,
\begin{equation}
-v_{M}\frac{n_{0}}{n}\left[p(n)-\frac{n'}{n}h(n)\right]=-[f'(n)+g'(n)]+\Omega^{2}\left(\frac{n}{n_{0}}-1\right)-l_{z}\frac{n}{n_{0}}\frac{\Omega^{2}}{v_{M}}q(n).\label{eq:diffeq}
\end{equation}
We note that the arbitrary functions $f,g,h,p,$ and $q$ can be written
entirely as functions of $n$.

\section{Soliton solutions}

In this section, we consider, in general, the electrons to be super
thermal governed by Lorentzian or kappa velocity distribution function
(VDF). The density of super thermal electrons with a kappa VDF can
be written as \cite{key-12,key-13,key-14,key-15},
\begin{equation}
n=n_{0}\left(1-\frac{2e\phi}{\kappa m_{e}\theta^{2}}\right)^{\nicefrac{1}{2}-\kappa},\label{eq:density}
\end{equation}
which reduces to the familiar Boltzmann relation in the limit $\kappa\rightarrow\infty$.
The temperature is represented by
\begin{equation}
\theta=\left[\left(\frac{2T_{e}}{m_{e}}\right)\frac{\kappa-3/2}{\kappa}\right]^{1/2}.\label{eq:theta}
\end{equation}
We note that validity of this particular VDF requires that the spectral
index $\kappa\geq3/2$. The expression for density {[}Eq.(\ref{eq:density}){]}
can be inverted to express the potential $\phi$ as,
\begin{equation}
\phi=\frac{1}{2}\frac{m_{e}}{e}\kappa\theta^{2}\left[1-\left(\frac{n}{n_{0}}\right)^{\frac{2}{1-2\kappa}}\right].\label{eq:phi}
\end{equation}
Using the above expression for $\phi$ in the expression for $g(n)$,
and carrying out the integration in Eq.(\ref{eq:vz}), we finally
write down the explicit nonlinear differential equation in $n$ from
Eq.(\ref{eq:diffeq}) as,
\begin{equation}
\alpha n''-\beta(n')^{2}+\xi=0,\label{eq:diffeq_final}
\end{equation}
where
\begin{eqnarray}
\alpha & = & 15nv_{M}^{2}\left[\sigma n^{8/3}-v_{M}^{2}+n^{2+\frac{2}{1-2\kappa}}\left(\frac{3-2\kappa}{1-2\kappa}\right)\right],\\
\beta & = & 5v_{M}^{2}\left[\sigma n^{8/3}-9v_{M^{2}}+3n^{2+\frac{2}{1-2\kappa}}\frac{(3-2\kappa)}{(1-2\kappa)^{2}}\right],\\
\xi & = & 3n^{4}\left[5v_{M}^{2}(1-n)-nl_{z}^{2}(5+3\sigma)+n^{2}l_{z}^{2}\left(3\sigma n^{2/3}+5n^{\frac{2}{1-2\kappa}}\right)\right].\label{eq:xi}
\end{eqnarray}
The above equations i.e. Eqs.(\ref{eq:diffeq_final}-\ref{eq:xi})
are written in terms of normalized variables. The ion density $n$
is normalized to its equilibrium density $n_{0}$, velocities to the
ion-sound velocity $ $$v_{s}=\sqrt{T_{e}/m_{i}}$, potential $\phi$
to $T_{e}/e$, and length to the ratio $v_{s}/\Omega$. The ratio
of the ion temperature to the electron temperature is denoted by $\sigma=T/T_{e}$.
Note that with this normalizations, the normalized ion velocity $v_{M}$
is the Mach number.

Equation (\ref{eq:diffeq_final}) describes the behavior of nonlinear
ion-acoustic wave including solutions of solitary waves. However,
in order to analytically isolate existence of solitary wave solutions,
we need to re-cast Eq.(\ref{eq:diffeq_final}) in a form,
\begin{equation}
\frac{1}{2}\left(\frac{dn}{d\eta}\right)^{2}+V(n)=0,\label{eq:sagdeev}
\end{equation}
where $V(n)$ is the pseudo or Sagdeev potential \cite{key-18}. A
potential-well structure of $V(n)$ ensures the existence of solutions
of solitary waves. The Eq.(\ref{eq:diffeq_final}), however, in its
present form with arbitrary $\kappa$ can not be reduced to a form
represented by Eq.(\ref{eq:sagdeev}), except in two limiting cases
$\kappa\rightarrow\infty$ and $\kappa\rightarrow3/2$. The first
is the well known case of Maxwellian electrons and the second is the
extreme limit of super thermal electrons. In what follows, we try
to use this two limiting cases.

\subsection{Limiting cases}

\subsubsection{Maxwellian electrons ($\kappa\rightarrow\infty$)}

\begin{figure}
\begin{centering}
\includegraphics[width=1\columnwidth]{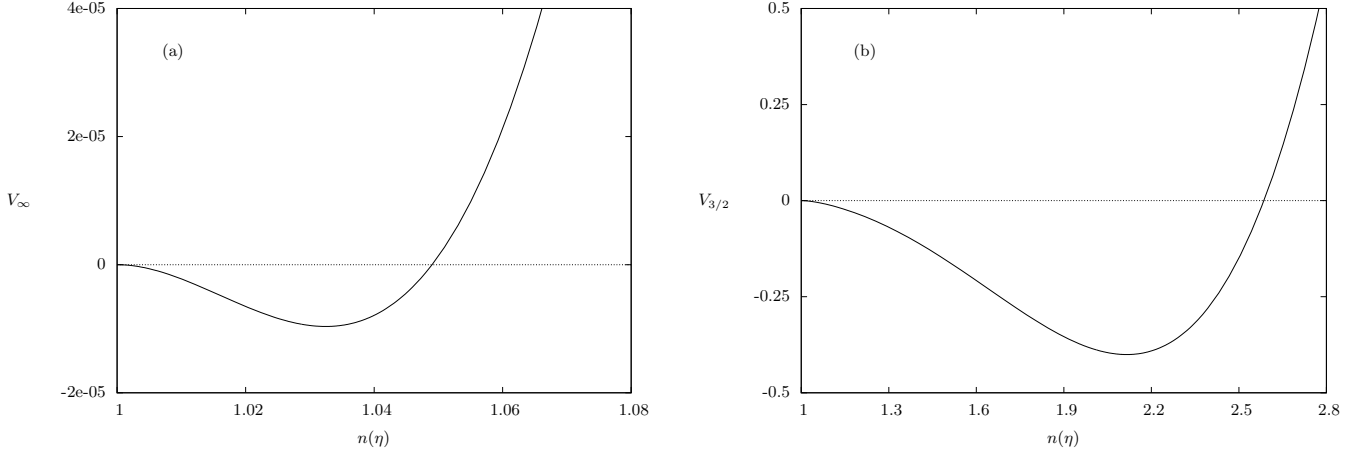}
\par\end{centering}

\caption{\label{fig:Sagdeev-potential}(a) Sagdeev potential for Maxwellian
electrons and (b) for highly super thermal electrons $(\kappa\sim3/2)$.
The parameters are $v_{M}=0.78,l_{z}=0.7,$ and $\sigma=0.2$. Note
the increase in depth and width of the potential in case of super
thermal electrons.}
\end{figure}

This is a well known case in which Eq.(\ref{eq:diffeq_final}) can
be reduced to the standard form of Eq.(\ref{eq:sagdeev}), with the
quasi potential $V_{\kappa}(n)$ as \cite{key-19},
\begin{eqnarray}
V_{\infty}(n) & = & [50v_{M}^{2}(\sigma n^{8/3}+n^{2}-v_{M}^{2})]^{-1}\times\left(n^{6}l_{z}^{2}[5(1-n)+3\sigma(1-n^{5/3})]^{2}\right.\nonumber \\
 &  & -\,5n^{5}v_{M}^{2}[\{10(1-n)(l_{z}^{2}-n)+n\sigma(9+6n^{5/3}-15n^{2/3})\nonumber \\
 &  & +\, l_{z}^{2}\sigma(6+9n^{5/3}-15n)\}\nonumber \\
 &  & \left.+\,10(l_{z}^{2}-1)n\ln n]+25n^{4}v_{M}^{4}(n-1)^{2})\right).\label{eq:sag-max}
\end{eqnarray}
In Fig.\ref{fig:Sagdeev-potential}(a), Sagdeev potential for Maxwellian
electrons is shown for a certain range of parameters. 

\subsubsection{Highly super thermal electrons ($\kappa\rightarrow3/2$)}

\begin{figure}
\begin{centering}
\includegraphics[width=1\columnwidth]{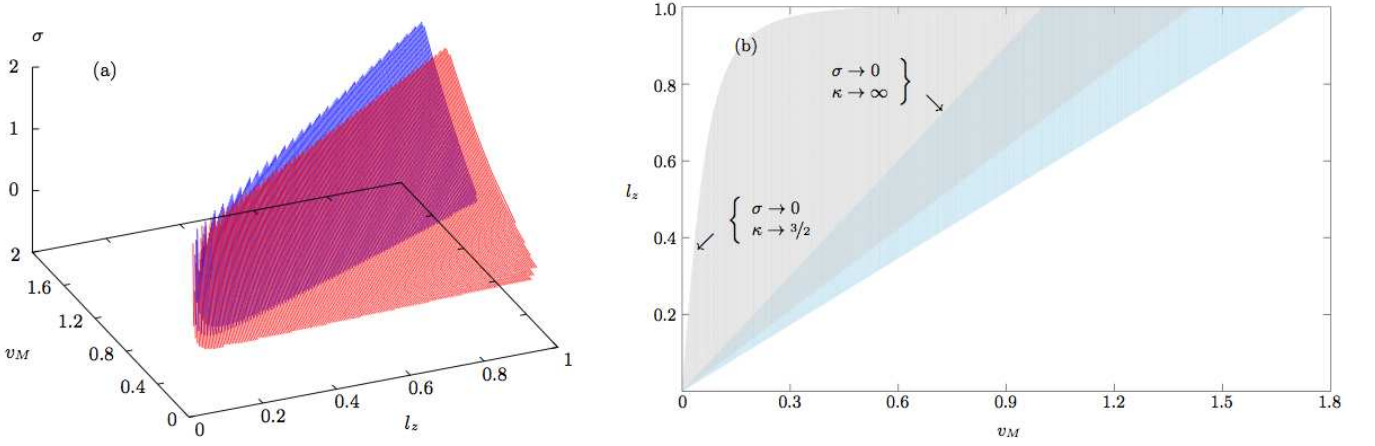}
\par\end{centering}

\caption{\label{fig:soliton-domain}(a) Domains of existence of solitons in
the two limiting cases in the $(v_{M},l_{z},\sigma)$ parameter space.
Solitons exist only left sides of the surfaces. The surface on the
front (red) is for $\kappa\sim3/2$ and the other (blue) is for $\kappa\rightarrow\infty$.
(b) Projection of the surfaces showing the regions of existence of
solitons in the $(v_{M},l_{z})$ space. The right side boundaries
of the regions are for higher $\sigma$.}
\end{figure}

We now carry out an asymptotic expansion of Eq.(\ref{eq:diffeq_final})
in the neighborhood of $\kappa\sim3/2$. After some simple and straight
forward but lengthy algebra, we can reduce Eq.(\ref{eq:diffeq_final})
to the following form,
\begin{equation}
\frac{d^{2}}{d\eta^{2}}G(n)=H(n),\label{eq:sag-kappa-1}
\end{equation}
where,
\begin{eqnarray}
G(n) & = & v_{M}^{2}\frac{1}{n^{2}}+(3-2\kappa)\frac{1}{n}+3\sigma n^{2/3},\label{eq:Gn}\\
H(n) & = & 2(n-1)+\left(\frac{l_{z}}{v_{M}}\right)^{2}\left[(3-2\kappa)n\,\ln n-\frac{6}{5}n(n^{5/3}-1)\right].
\end{eqnarray}
Multiplying both sides of Eq.(\ref{eq:sag-kappa-1}) by $dG(n)/d\eta$
and integrating, we to get,
\begin{equation}
\frac{1}{2}\left(\frac{d}{d\eta}G(n)\right)^{2}=\int H(n)\,\frac{d}{d\eta}G(n)\, d\eta+C_{2},
\end{equation}
from which, we can finally write the standard form for the quasi potential
equation in the form of Eq.(\ref{eq:sagdeev}) with the quasi or Sagdeev
potential as,
\begin{eqnarray}
V_{\nicefrac{3}{2}}(n) & = & [50v_{M}^{2}\{2v_{M}^{2}+n(3-2\kappa-2n^{5/3}\sigma)\}]^{-1}\times\left(n^{6}l_{z}^{2}[6\sigma(n^{5/3}-1)\right.\nonumber \\
 &  & -\,5\ln n(3-2\kappa)]^{2}+100n^{4}v_{M}^{4}(n-1)^{2}\nonumber \\
 &  & +\, n^{4}v_{M}^{2}[100n(3-2\kappa)\{(l_{z}^{2}-1)(n-1)-(l_{z}^{2}-n)\ln n\}\nonumber \\
 &  & \left.-\,60n\sigma\{3n+n^{5/3}(2n-5)+l_{z}^{2}(2-5n+3n^{5/3})\}]\right).\label{eq:sag-kappa}
\end{eqnarray}
The constant of integration $C_{2}$ is evaluated by demanding the
boundary condition $V(n)|_{n=1}=0$. The Sagdeev potential for this
limiting case is shown in Fig.\ref{fig:Sagdeev-potential}(b). In
Fig.\ref{fig:soliton-domain}(a), we have shown the domain of existence
of a soliton in the $(v_{M}-l_{z}-\sigma)$ parameter space for both
these limiting cases. A soliton, if any, can exist only left of the
surfaces (please see the captions for details). The projection of
the domains in the $(v_{M}-l_{z})$ space is shown in Fig.\ref{fig:soliton-domain}(b).
The left side of the regions are limited by $\sigma\rightarrow0$.
In the overlapping region, solitons in both extreme cases can form. 

\subsubsection{Limiting Mach numbers}

\begin{figure}
\begin{centering}
\includegraphics[width=1\columnwidth]{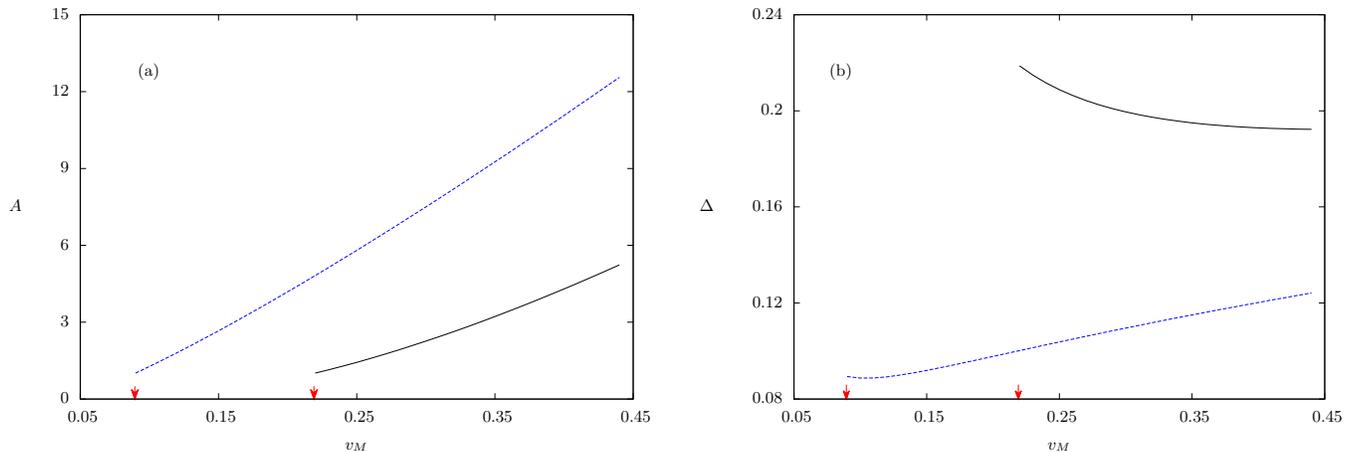}
\par\end{centering}

\caption{\label{fig:limiting -mach}(a) Dependences of soliton amplitude $A$
on Mach number. (b) Dependences of soliton width $\Delta$ on Mach
number. The arrows show the position of lower limits of the respective
Mach numbers. The blue (dashed) curves are for $\kappa\sim3/2$ and
the solid (black) curves are for $\kappa\rightarrow\infty$ (Maxwellian
electrons). The parameters are same as in Fig.\ref{fig:Sagdeev-potential}.}
\end{figure}

The limiting Mach numbers can be found out by imposing the condition
that $V(n)$ must have a local maximum at $n=1$, which translates
to the conditions
\begin{equation}
l_{z}\sqrt{1+\sigma}<v_{M}<\sqrt{1+\sigma},\label{eq:limit_max}
\end{equation}
for Maxwellian electrons and
\begin{equation}
l_{z}\sqrt{\kappa-3/2+\sigma}<v_{M}\sqrt{\kappa-3/2+\sigma}.\label{eq:limit_kappa}
\end{equation}
for highly super thermal electrons ($\kappa\sim3/2$). The latter
condition reduces to $l_{z}\sqrt{\sigma}<v_{M}<\sqrt{\sigma}$ for
$\kappa\rightarrow3/2$ . So, with highly Lorentzian electrons, solitons
can form even with lower values of Mach numbers. In Fig.\ref{fig:limiting -mach},
we have shown the dependence of soliton amplitude $A$ and width $\Delta$,
respectively, with the Mach number for both these cases. Note that
the amplitude $A$ is determined by first zero of the Sagdeev potential
away from $n=1$ and the soliton width $\Delta=A/\sqrt{d}$, where
$d$ is the maximum depth of the Sagdeev potential determined by the
condition $dV(n)/dn|_{n\neq1}=0$. As can be seen from Fig.\ref{fig:limiting -mach},
in case of highly super-thermal electrons, the soliton can be very
steep (smaller width) and can reach high amplitude in comparison with
thermal electrons.

Note that the range of the valid Mach number is exclusively dictated
by the direction of propagation of the soliton with the higher limit
parallel to the magnetic field ($l_{z}\rightarrow1$).

\subsection{Arbitrary $\kappa$ --- \emph{numerical solution}}

\begin{figure}
\begin{centering}
\includegraphics[width=1\columnwidth]{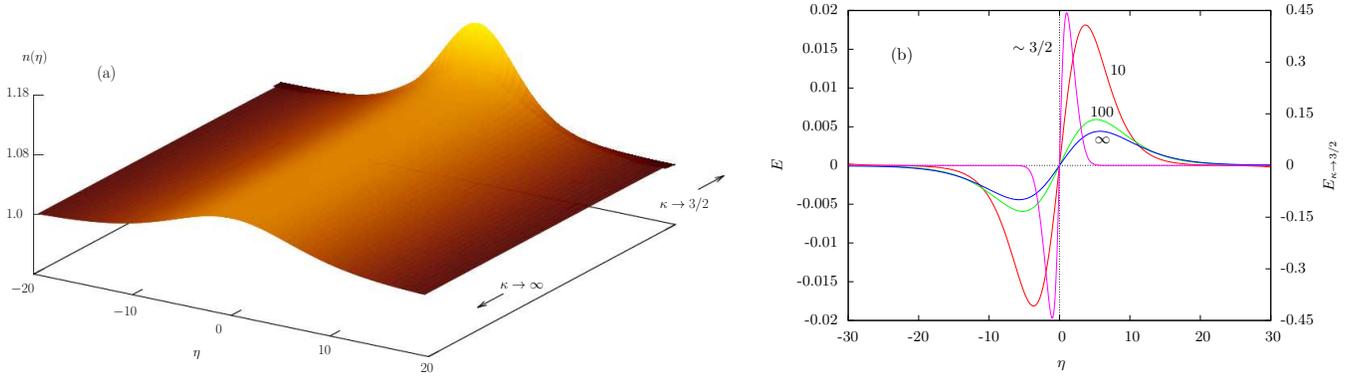}
\par\end{centering}

\caption{\label{fig:soliton-arb-kappa}(a) Evolution of soliton as $\kappa$
approaches $3/2$ from a large value. (b) The bi-polar electric field
corresponding to (a). Use the right hand side axis for the case when
$\kappa\sim3/2$, which is very large compared to others. }
\end{figure}

As mentioned earlier, Eq.(\ref{eq:diffeq_final}) can not be reduced
to standard form for arbitrary $\kappa$, which can analytically demonstrate
the existence of a suitable quasi potential and thus solitary wave
solutions, which \emph{however}, does not rule out existence of soliton-like
solutions for arbitrary $\kappa$. In order to solve the second order,
nonlinear differential equation, Eq.(\ref{eq:diffeq_final}) numerically
for solitary wave solutions, we use the shooting method, imposing
the boundary conditions $n(\eta)|_{\eta\rightarrow\pm M}=1$ and start
with an initial condition $n'(\eta)|_{\eta=0}=0$, where $M$ is a
large number. In practice, we need to solve only one half on the $\eta$-axis
in the range $\eta\in[0,+M]$ as the the solutions $n(\eta)$ are
always symmetric in the range $[-M,+M]$, which can be easily seen
from the invariance of Eq.(\ref{eq:diffeq_final}) for $\eta\rightarrow-\eta$.
In Fig.\ref{fig:soliton-arb-kappa}(a), we have shown the evolution
of the soliton as obtained from numerical solution of Eq.(\ref{eq:diffeq_final})
for arbitrary $\kappa$. The corresponding bi-polar electric field
representing the solitary density structure is shown in Fig.\ref{fig:soliton-arb-kappa}(b).
Note that the corresponding electric field for a density soliton can
be found from the relation,
\begin{equation}
E(\eta)=-\frac{1}{n}\,\frac{dn}{d\eta}.\label{eq:electric_field}
\end{equation}

\begin{figure}
\begin{centering}
\includegraphics[width=1\columnwidth]{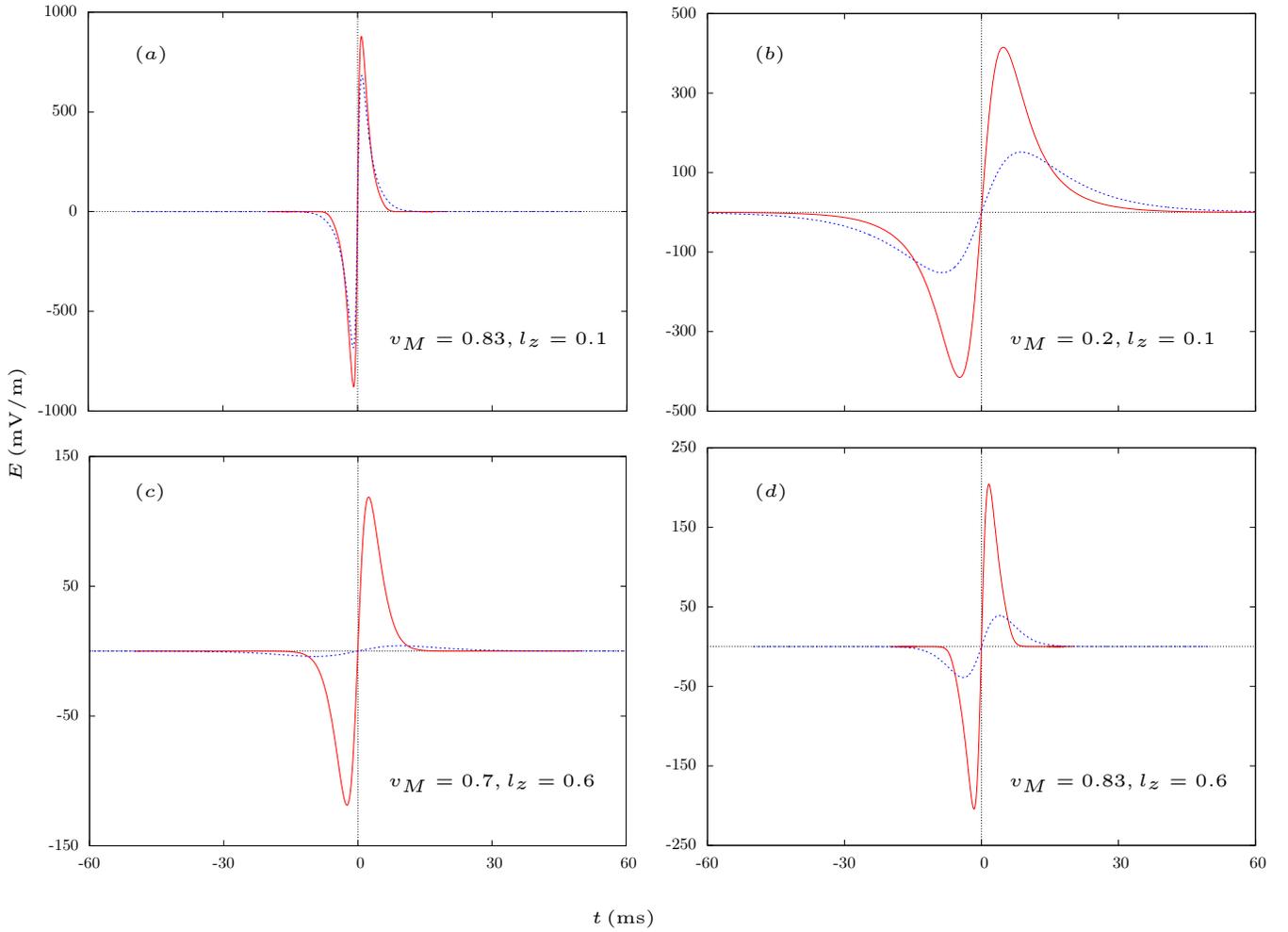}
\par\end{centering}

\caption{\label{fig:Bi-polar-electrostatic-pulse}Bi-polar electrostatic pulse
in real-time units. The solid (red) curves are for highly Lorentzian
electrons ($\kappa\sim3/2$) and the dashed (blue) curves are for
Maxwellian electrons. }
\end{figure}

We note that in the auroral region, the ion to electron temperature
ratio $\sigma\sim0.25$ and the ion gyro frequency $(\Omega/2\pi)\sim100\,{\rm Hz}$
\cite{key-22}. Consider now various bi-polar electrostatic structures
observed in different regions viz. auroral region of the ionosphere
and earth's magnetosphere. These structures show peak-to-peak variation
of electric field ranging from few hundreds of micro volts per meter
to milli volts per meter within an interval of micro seconds to milli
seconds. Usually the amplitudes of pulses in the auroral region are
larger (milli volts per meter) having a larger duration of the order
of milli seconds \cite{key-7}. Those observed in the magnetosheath
region are of much smaller amplitudes (micro volts per meter) with
a very short duration (micro seconds) \cite{key-21}. Note that the
bi-polar structures represented in Fig.\ref{fig:soliton-arb-kappa}(b)
are in the rest frame of the solitons, which can be transformed to
the duration of the pulses in the rest frame of the detector (on board
the satellite). In Fig.\ref{fig:Bi-polar-electrostatic-pulse}, we
show various bi-polar electrostatic structures for these parameters
in real-time units. In all panels of Fig.\ref{fig:Bi-polar-electrostatic-pulse},
the pulse is of shorter duration and larger height for highly Lorentzian
electrons than Maxwellian electrons. The peak-to-peak electric field
variation due to Lorentzian electrons in panel (d) is $\sim250\,{\rm mV}/{\rm m}$
with a duration $\sim20\,{\rm ms}$, which is typical in these regions
\cite{key-7,key-22}. For the same parameters, the peak-to-peak variation
is much smaller $\sim90\,{\rm mV}/{\rm m}$ with a longer duration
$\sim40\,{\rm ms}$ for Maxwellian electrons. The pulses due to Maxwellian
electrons are comparable to Lorentzian electrons in both pulse height
and width only when the pulses become very large, which are however,
not observed in these regions of space plasma. So, we conclude that
the deviation of electron temperature from thermal behavior is an
essential factor, which needs to be taken into account in order to
explain the relatively small-amplitude and steep (smaller width) bi-polar
electrostatic structures in the auroral regions.

We further note that the effect due to super-thermal electrons is
more pronounced when $l_{z}\sim1$ or $\theta\sim0^{\circ}$, where
$\theta$ defines the angle between the ambient magnetic field and
the direction of propation of the soliton. This is consistent with
observvation of these solitons in magnetospheric plasmas\cite{key-2,key-3,key-4,key-5}.

\section{Two-temperature electrons : \emph{small amplitude solitons}}

 In this section, we consider the  physical scenario with two-temperature
electrons, which is motivated by the experimental observations such
as those by the Cluster spacecrafts \cite{key-2}. We consider two
species of electrons --- hot and cold one. We still assume that both
species of electrons are Lorentzian, which gives us freedom to investigate
the situation both for thermal and super-thermal cases. All mathematical
analyses remaining same, differing only in the expression denoting
the combined density of the electrons {[}see Eq.(\ref{eq:density}){]},
\begin{equation}
n_{e}=n_{c}+n_{h}=n_{c0}\left(1-\frac{2e\phi}{\kappa m_{e}\theta_{c}^{2}}\right)^{\nicefrac{1}{2}-\kappa}+n_{h0}\left(1-\frac{2e\phi}{\kappa m_{e}\theta_{h}^{2}}\right)^{\nicefrac{1}{2}-\kappa},\label{eq:two-density}
\end{equation}
and the quasi-neutrality condition {[}see Eq.(\ref{eq:boltzmann}){]},
\begin{equation}
n=n_{c}+n_{h},\label{eq:quasi}
\end{equation}
where the subscripts `$c$' and `$h$' refer to the \emph{cold} and
\emph{hot} electrons with $\theta_{c,h}$ respectively being functions
of $T_{c,h}$, the temperatures of cold and hot electrons. The primary
drawback in this formalism that we can not find an equivalent Sagdeev
potential for solitons of arbitrary amplitudes as Eq.(\ref{eq:two-density})
can not be inverted to find a unique analytical expression for the
potential $\phi$ in terms of the densities. In what follows, we shall
expand Eq.(\ref{eq:two-density}) around $\phi=0$, assuming the analysis
for only small-amplitude solitons, so that we can find out the Sagdeev
potential which, however, \emph{will be valid} only in the limit of
small amplitude.

A Taylor expansion of Eq.(\ref{eq:two-density}) around $\phi=0$,
to the zeroth order enables us to write the plasma potential (un-normalized)
as,
\begin{equation}
\phi\simeq\frac{T_{c}T_{h}}{e}\left(\frac{n-n_{c0}-n_{h0}}{n_{c0}T_{h}+n_{h0}T_{c}}\right)\left(\frac{2\kappa-3}{2\kappa-1}\right).\label{eq:two-phi}
\end{equation}
As before, we normalize the variables and write the normalized electron
densities as $\nu_{c,h}=n_{c0,h0}/n_{0}$ with the condition $v_{c}+v_{h}=1$.
The temperature ratio of the cold and hot electrons is $\beta=T_{c}/T_{h}$
and $\sigma=T/T_{h}$. Proceeding as before, we can reduce the equations
to the form given by Eq.(\ref{eq:sagdeev}) and write the Sagdeev
potential (for small amplitude, $n\rightarrow1$) as 
\begin{equation}
V_{n\rightarrow1}\simeq\frac{n^{4}[100f_{1}^{2}v_{M}^{4}\delta n^{2}+\delta n_{1}^{2}-20nf_{1}v_{M}^{2}\{5f_{2}(n+l_{z}^{2})\delta n^{2}+\delta n_{2}\}]}{200v_{M}^{2}(f_{1}\sigma n^{8/3}-f_{1}v_{M}^{2}+f_{2}n^{3})},\label{eq:small-sagdeev}
\end{equation}
where $\delta n=n-1$ and other quantities are given by,
\begin{equation}
\left.\begin{array}{rcl}
f_{1} & = & (2\kappa-1)\,\{1-\nu_{h}(1-\beta)\},\\
\\
f_{2} & = & (2\kappa-3)\beta,\\
\\
\delta n_{1} & = & 5f_{2}l_{z}(n^{2}-1)+6f_{1}l_{z}\sigma(n^{5/3}-1),\\
\\
\delta n_{2} & = & 3f_{1}\sigma[3n+n^{5/3}(2n-5)+l_{z}^{2}(3n^{5/3}-5n+2).
\end{array}\right\} \label{eq:var-1}
\end{equation}
The quantities $\delta n,\delta n_{1,2}\rightarrow0$ as $n\rightarrow1$
and it is easy to see that the function $V_{n\rightarrow1}$ satisfies
all the requirements of Sagdeev potential. However, its interpretation
is valid only in the limit $n\rightarrow1$.

To complete the analysis, we now take the small amplitude limit of
the $V_{n\rightarrow1}$ in Eq.(\ref{eq:sagdeev}) by expanding around
$n=1$. The resultant equation, to the third order, is,
\begin{equation}
\frac{1}{2}\left(\frac{d}{d\eta}\delta n\right)^{2}+\chi_{1}\,\delta n^{2}+\chi_{2}\,\delta n^{3}=0,\label{eq:kdv}
\end{equation}
which we recognize as the Korteweg de-Vries (K-dV) equation in the
transformed space-time co-ordinate $\eta$, describing soliton solutions.
The factors $\chi_{1}$ and $\chi_{2}$ are given by

\begin{eqnarray}
\chi_{1} & = & \frac{f_{1}v_{M}^{2}-f_{3}l_{z}^{2}}{2v_{M}^{2}f_{4}},\label{eq:x1}\\
\chi_{2} & = & \frac{1}{18v_{M}^{2}f_{4}^{2}}\left[3f_{1}f_{2}l_{z}^{2}(7\sigma-15v_{M}^{2})+9l_{z}^{2}f_{2}^{2}\right.\nonumber \\
 &  & \left.+\,4f_{1}^{2}\{9v_{M}^{4}+3l_{z}^{2}\sigma^{2}-v_{M}^{2}\sigma(1+11l_{z}^{2})\}\right]\label{eq:x2}
\end{eqnarray}
 and 
\begin{equation}
\left.\begin{array}{rcl}
f_{3} & = & f_{2}+f_{1}\sigma,\\
\\
f_{4} & = & f_{1}v_{M}^{2}-f_{3}.
\end{array}\right\} \label{eq:var-2}
\end{equation}
In the above equations, we note that the factors $f_{1,2,3}\geq0$.
The amplitude $A$ and width $\Delta$ of this K-dV soliton is given
by,
\begin{eqnarray}
A & = & -\frac{\chi_{1}}{\chi_{2}},\label{eq:a}\\
\Delta & = & \sqrt{-\frac{2}{\chi_{1}}},\label{eq:delta}
\end{eqnarray}
so that for existence of a soliton, $\chi_{1}<0$, which translates
to the condition,
\begin{equation}
l_{z}\sqrt{f_{\kappa}+\sigma}<v_{M}<\sqrt{f_{\kappa}+\sigma},\label{eq:soliton-cond}
\end{equation}
where the factor
\begin{equation}
f_{\kappa}=\frac{f_{3}}{f_{1}}=\frac{\beta(2\kappa-3)}{(2\kappa-1)\,\{1-\nu_{h}(1-\beta)\}},\label{eq:f}
\end{equation}
represents the effect of two-temperature electrons. Under various
limiting cases, we can reduce this condition to simple forms,
\begin{eqnarray}
l_{z}\sqrt{\frac{\beta}{1-\nu_{h}(1-\beta)}+\sigma} & < & v_{M}<\sqrt{\frac{\beta}{1-\nu_{h}(1-\beta)}+\sigma},\qquad\kappa\rightarrow\infty\label{eq:cond-max}\\
l_{z}\sqrt{\frac{\beta(\kappa-3/2)}{1-\nu_{h}(1-\beta)}+\sigma} & < & v_{M}<\sqrt{\frac{\beta(\kappa-3/2)}{1-\nu_{h}(1-\beta)}+\sigma},\qquad\kappa\rightarrow3/2\label{eq:cond-kappa}
\end{eqnarray}
which further reduce to (\ref{eq:limit_max}) and (\ref{eq:limit_kappa})
if we consider only single temperature for electrons ($\nu_{h}\rightarrow1$).
As the factor $f_{\kappa}$ lies between zero and unity, the effect
of two different temperatures is not very significant expect when
$\sigma$ is small. We however note that the parameters $\beta$ and
$\nu_{h}$ related to two different temperatures can dictate the exact
parameter regime, where solitons may exist.

\section{Conclusions}

In this paper, we have discussed the physical situation where electrostatic
solitary structures can form in a warm plasma immersed in a constant
background magnetic field and in the presence of a non-thermal electron
fluid. The results are viewed in reference to the bi-polar ESWs, typically
observed in the auroral regions. We have shown that in presence of
a super-thermal electron population, large amplitude solitary structure
can form in the parallel direction, which otherwise is not possible
for Maxwellian electrons. We have also emphasised that with a super-thermal
electron fluid, solitons can form for lower Mach numbers. We have
presented our results in the real-time units of the bi-polar ESWs,
which are basically manifestations of solitary waves, and our results
are in good agreement to the available experimental data in the auroral
region.

We have further considered the presence of two components of electron
populations, both of which can be super-thermal, in view of the observations
by the certain space-borne experiments viz. Cluster spacecrafts \cite{key-21}.
However we find only marginal modifications to the ESWs by including
the two-temperature electrons, and conclude that they may have only
minor role in dictating the exact parameter details of the ESWs.


\begin{thebibliography}{10}
\bibitem{key-21}J. S. Pickett et. al., Adv. Space Res. 41, 1666 (2008).

\bibitem{key-2-1}B. T. Tsurutani, J. K. Arballo, G. S. Lakhina, C.
M. Ho, B. Buti, J. S. Pickett, and D. A. Gurnett, Geophys. Res. Lett.
25, 4117 (1998).

\bibitem{key-3-1}G. S. Lakhina, B. T. Tsurutani, H. Kojima, and H.
Matsumoto, J. Geophys. Res. 105, 27791 (2000).

\bibitem{key-4-1}J. R. Franz, P. M. Kintner, J. S. Pickett, and L.-J.
Chen, J. Geophys. Res. 110 A09212 (2005).

\bibitem{key-5-1}L.-J. Chen, J. S. Pickett, P. Kintner, J. Franz,
and D. Gurnett, J. Geophys. Res. 110 A09211 (2005).

\bibitem{key-6-1}J. D. Williams, L.-J. Chen, W. S. Kurth, D. A. Gurnett,
M. K. Dougherty, and A. M. Rymer, Geophys. Res. Lett. 32 L17103 (2005).

\bibitem{key-7-1}J. D. Williams, L.-J. Chen, W. S. Kurth, D. A. Gurnett,
and M. K. Dougherty, Geophys. Res. Lett. 33, L06103 (2006).

\bibitem{key-9-1}M. P. Bora, B. Choudhury, and G. C. Das, Astrophys.
Sp. Sc. (2012).

\bibitem{key-1}P. Meuris and F. Verheest, Phys. Lett. A 219, 2992
(1996).

\bibitem{key-2}J. S. Pickett et. al., Ann. Geophys. 22, 2515 (2004).

\bibitem{key-3}G. T. Marklund et. al., Nonlin. Processes Geophys.
11, 709 (2004).

\bibitem{key-4}M. Temerin, K. Cerny, W. Lotko, and F. S. Mozer, Phys.
Rev. Lett. 48, 1175 (1982).

\bibitem{key-5}R. Bostr\"om et. al., Phys. Rev. Lett. 61, 82 (1988).

\bibitem{key-6}F. S. Mozer et. al., Phys. Rev. Lett. 79, 1281 (1997).

\bibitem{key-7}S. Bounds et. al., J. Geophys. Res. 104, 28709 (1999).

\bibitem{key-8}J. Dombeck et. al., J. Geophys. Res. 106, 19013 (2001).

\bibitem{key-9}P. Dovner, A. Eriksson, R. Bostr\"om, and B. Holback,
Geophys. Res. Lett. 21, 1827, (1994).

\bibitem{key-10}Y. Chen, Z.-Y. Li, W. Liu, and Z.-D. Shi, Phys. Plasmas
7, 371 (2001).

\bibitem{key-11}E. Marsch et. al., J. Geophys. Res. 87 (A1), 52 (1982).

\bibitem{key-12}C.-Y. Ma and D. Summers, Geophys. Res. Lett. 25,
4099 (1998).

\bibitem{key-13}D. Summers, S. Xue, and R. M. Thorne, Phys. Plasmas
1, 6 (1994).

\bibitem{key-14}D. Summers and R. M. Thorne, Phys. Fluids B 3, 1835
(1991).

\bibitem{key-15}D. Summers and R. M. Thorne, J. Geophys. Res. 97
(A11), 16827 (1992).

\bibitem{key-16}M. N. S. Qureshi et al., Proc. Tenth Int. Solar Wind
Conf. p 489 (2003).

\bibitem{key-17}O. A. Pokhotelov, O. G. Onishchenko , M. A. Balikhin,
L. Stenflo, and P. K.Shukla, J. Plasma Phys. 73, 981 (2007).

\bibitem{key-18}R. Z. Sagdeev, Rev. Plasma Phys. 4, 23 (1966).

\bibitem{key-19}M. K. Kalita and S. Bujarbarua, J. Phys. A 16, 439
(1983).

\bibitem{key-22}J. Shi et. al., Ann. Geophys. 26, 1431 (2008).\end{thebibliography}
\end{document}